# The Sublimative Evolution of (486958) Arrokoth

Jordan K. Steckloff[1,2]
Carey M. Lisse[3]
Taylor K. Safrit[4]
Amanda S. Bosh[4]
Wladimir Lyra[5]
Gal Sarid[6]

[1]Planetary Science Institute, Tucson AZ
[2]University of Texas at Austin, Austin, TX
[3]Johns Hopkins University, Applied Physics Laboratory, Laurel, MD
[4]Massachusetts Institute of Technology, Cambridge, MA
[5]New Mexico State University, Las Cruces, NM
[6]SETI Institute, Mountain View, CA

## Abstract

We consider the history of *New Horizons* target (486958) Arrokoth in the context of its sublimative evolution. Shortly after the Sun's protoplanetary disk (PPD) cleared, the newly intense sunlight sparked a sublimative period in Arrokoth's early history that lasted for ~10 – 100 Myr. Although this sublimation was too weak to significantly alter Arrokoth's spin state, it could drive mass transport around the surface significant enough to erase topographic features on length scales of ~10 - 100 m. This includes craters up to ~50 – 500 m in diameter, which suggests that the majority of Arrokoth's craters may not be primordial (dating from the merger of Arrokoth's lobes), but rather could date from after the end of this sublimative period. Thereafter, Arrokoth entered a Quiescent Period (which lasts to the present day), in which volatile production rates are at least 13 orders of magnitude less than the ~$10^{24}$ molecules/s detection limit of the *New Horizons* spacecraft (*Lisse et al. 2020*). This is insufficient to drive either mass transport or sublimative torques. These results suggest that the observed surface of Arrokoth is not primordial, but rather dates from the Quiescent Period. By contrast, the inability of sublimative torques to meaningfully alter Arrokoth's rotation state suggests that its shape is indeed primordial, and its observed rotation is representative of its spin state after formation.

## I.        Introduction

(486958) Arrokoth is a cold classical Kuiper Belt Object (KBO), which is thought to have formed at 44.6 au from the Sun, where it has resided ever since (*Stern et al. 2019*). This stability is unique amongst all objects previously visited by spacecraft, which are thought to have moved and migrated due to instabilities among the giant planets and disk-planet interaction after formation. Thus, Arrokoth provides an opportunity to study how thermally driven processes could have evolved this object over the age of the Solar System, from formation in the Sun's protoplanetary disk (PPD) to the present.  Because Arrokoth is rich in  ices (*Grundy et al. 2020; Lisse et al. 2020*), early sublimative processes could have dramatically altered its surface features and appearance, and even its behavior if vigorous enough, e.g.  affecting the spin state of the body via the effects of sublimative torques.  Indeed, such processes could be quite important in the evolution of Arrokoth, which (i) is thought to have formed from a gentle merger of two separate objects that bled off significant orbital and rotational angular momentum (*Hirabayashi et al. 2020, McKinnon et al. 2020*) and (ii) shows sufficient crater counts to date its surfaces to the earliest part of the Solar System's history (*Spencer et al. 2020*). Here we explore how sublimative activity may have affected Arrokoth over its history.

## II.        Periods of Arrokoth's History

Arrokoth formed in extremely cold conditions in the Sun's protoplanetary disk (PPD), sufficient for CO (*Krijt et al. 2018*) and other hypervolatile species (such as $N_2$, and $CH_4$) to condense into solids. Later, the Sun's T Tauri winds and gas accretion cleared out the protoplanetary disk (PPD); Arrokoth's surface was newly exposed to solar heating. At this point, the surface of Arrokoth was likely pristine, with hypervolatile CO, $N_2$, and $CH_4$ present (*Lisse et al. 2020)*.  The sublimation rate of these ices would have been very high, representing a maximum sublimative flux from the surface. However, as these volatiles sublimated away and depleted the body, Arrokoth would eventually cease to vigorously outgas, and would enter a steady state sublimative regime that exists to the present day.   For this reason, we must separately consider these two periods: the "Sublimative Period" (shortly after the clearing of the Sun's PPD), and the "Quiescent Period" (began once Arrokoth reached a thermodynamic steady state).

*Arrokoth's Sublimative Period*

We can estimate the duration of the Sublimative Period by computing how long the post PPD thermal wave would require to reach the center of the body (and hence the time after which the primitive thermal structure is lost). Although there has been some advancement in past work regarding the thermophysical evolution of KBO surfaces and interiors, there are still many basic open questions, large parameter uncertainties, and a lack of constraints for ranges of physical parameters. Specific to our modeling work in this paper are the important issues of object size, density, and assumed composition. In many respects, while Arrokoth is clearly a member of the cold-classical KBO dynamical grouping, its size and probable density of $\sim 300 - 500$ kg/m$^3$ (*Hirabayashi et al. 2020; McKinnon et al. 2020*) is physically most akin to a large comet nucleus (e.g., *Richardson et al. 2007; Thomas et al. 2013; Patzold et al. 2016, Kokotanekova et al. 2017; Pruesker et al. 2017*). The two distinct lobes may even be considered as proxies for two small comet nuclei (roughly 7.5 and 10 km in radius), as their internal thermal evolution has probably initiated and progressed with minimal cross-influence in the deep interior, similarly to other small icy bodies (*e.g., Davidsson et al. 2016, McKinnon et al. 2020*).

In contrast, previous work dealing with the thermophysical evolution of KBOs has largely focused on significantly larger objects with higher densities than Arrokoth (*e.g., De Sanctis et al., 2001, 2007; Merk & Prialnik, 2003; Choi et al, 2002; Prialnik et al. 2008; Sarid & Prialnik, 2009; Desch et al. 2009; Guilbert-Lepoutre et al. 2011*). Although some of the studied examples are only an order of magnitude greater in volume than Arrokoth (*e.g., De Sanctis et al. 2001; Prialnik et al. 2008; Sarid & Prialnik, 2009*), these examples are nevertheless much too large to be applicable to Arrokoth; especially in the effects of radiogenic heating, for which the amplitude of heating depends directly on size. The net thermal energy deposition results form a balance between heat loss through the surface and heat production within the interior. Thus, the smaller the object's radius, the lower the maximum temperature it can attain. Indeed, while sophisticated models exhibit the dependence on size, density, composition, and radionuclide abundance, many studies of icy body radiogenic heating have shown that the amplitude of this effect is negligible for water ice rich bodies (mostly in mixed crystalline and amorphous form) with radii smaller than $\sim$10-15 km (*e.g., Prialnik & Podolak 1995; De Sanctis et al. 2001; Choi et al. 2002; Prialnik et al. 2008, Sarid & Prialnik 2009*).

To estimate the characteristic timescale of the Sublimative Period, we rearrange the thermal skin depth equation, which measures the distance below the surface at which the amplitude of a surface thermal wave decreases by $1/e$:

$$d_{skin} = \sqrt{\mathcal{H}t} \qquad (1)$$

$$t = \frac{d_{skin}^2}{\mathcal{H}} \qquad (2)$$

where $d_{skin}$ is the thermal skin depth, $\mathcal{H}$ is the thermal diffusivity, and $t$ is the elapsed time. Thermal diffusivity measures the ease with which a heat wave diffuses into a surface, and is related to the thermal conductivity ($k$), specific heat capacity ($c_p$), and density ($\rho$) by

$$\mathcal{H} = \frac{k}{\rho c_p}. \qquad (3)$$

For Arrokoth, we use its most plausible density range of ~300 - 500 kg/m³ (*Hirabayashi et al. 2020; McKinnon et al. 2020*), and the specific heat capacities for volatile ices, which are on the order of ~2000 J/kg K (the specific heat capacity of water ice). The thermal conductivity of porous cometary material is on the order of ~$10^{-2} - 10^{-1}$ W/K m (*Gundlach & Blum, 2012*), resulting in a thermal diffusivity of ~$10^{-8} - 10^{-7}$ m²/s.

Alternatively, thermal inertia ($I$), which measures a surface's resistance to changing its temperature in response to a changing thermal environment, is related to thermal conductivity, specific heat capacity, and density by

$$I = \sqrt{k\rho c_p} \qquad (4)$$

thus

$$\mathcal{H} = \frac{I^2}{\rho^2 c_p^2}. \qquad (5)$$

The thermal diffusivity of the surface of Comet 67P/Churyumov-Gerasimenko is $80^{+80}_{-40}$ JK⁻¹m⁻²s⁻⁰·⁵ (*Marshall et al. 2018*) with a density of $537.8 \pm 0.7$ kg/m³ (*Pruesker et al. 2017*), with a specific heat capacity modeled to be around ~350 (*Grundy et al. 2020*) to ~1000 J kg⁻¹K⁻¹ (*Hu et al. 2017*), $2 \times 10^{-8}$ ($6 \times 10^{-9}$ - $9 \times 10^{-8}$) m²/s, which is consistent with this range of thermal diffusivities.

These values are smaller than the thermal diffusivities of typical geological materials (~$10^{-6}$ $\frac{m^2}{s}$; *Drury, 1987*) due to the low densities and high porosities of small icy bodies[1]. From this thermal diffusivity range, the thermal wave of the Sun would reach the centers of Arrokoth's lobes ~10 − 100 Myr after the dispersal of the PPD. This calculation assumes that the clearing of the PPD occurred instantaneously. While this is not strictly true (and indeed was likely to have occurred over a timescale of up to a few million years; *Williams & Cieza, 2011*), this is sufficiently short to not affect these order of magnitude estimates of Arrokoth's Sublimative Period duration.

---

[1] Grundy et al. (2020) assumed thermal properties that recreate ground-based observations of KBOs, but result in thermal diffusivity values 3-4 orders of magnitude small still, which would result in the Sublimative Period lasting for ~100 Gyr

It is important to remember that the two lobes of Arrokoth have a smaller effective radius than their long-axis radius. When considering a more realistic representation of the "flat ellipsoid" shape of Arrokoth's lobes (*McKinnon et al. 2020*), we can expect the internal thermal evolution to be less secular and advance in a more episodic pattern and at a faster rate at different locations within the body, especially if internal inhomogeneities are present (*e.g. Rosenberg & Prialnik 2010, Guilbert-Lepoutre et al. 2016*). Hence, these thermal evolution timescales are likely an estimate for the upper limit of the volatile-active, sublimative period, in Arrokoth's evolution.

During the Sublimative Period, some volatiles outgassed vigorously, while others maintained extremely low production rates. To compute the sublimative flux of these volatiles, we use the model of *Steckloff et al. (2015)*, which solves the energy balance equation at the surface of the body, accounting for solar heating, and both radiative and sublimative cooling. This model uses first principles (Clausius-Clapyron relation, Langmuir-Knudsen equation of sublimation rates [*Langmuir, 1913*], conservation of energy, and conservation of momentum) to solve for the sublimative molecular flux and sublimative back reaction pressure exerted on the surface. Using this model, we can constrain the sublimative flux of CO from the surface of Arrokoth to up to ~4 x $10^{29}$ molecules per second. $N_2$ ice would behave similarly to CO (*Lisse et al. 2020*). Due to their much higher enthalpy of sublimation, sublimative cooling from other common icy volatiles such as $H_2O$ and $CO_2$ are at least eighteen orders of magnitude lower at 44.6 AU, Arrokoth's distance from the Sun (see figure 1), and are therefore insignificant in the Sublimative Period. Thus, Arrokoth would effectively cease to outgas vigorously once depleted of CO and other minor supervolatiles.

*Arrokoth's Quiescent Period*

Once the thermal wave and supervolatile sublimation fronts reach Arrokoth's center, the supervolatiles will be depleted, ending Arrokoth's Sublimative Period. More refractory volatiles (e.g., $CO_2$, $H_2O$) would still be present within Arrokoth, however temperatures would be much too low to allow these volatiles to sublimate vigorously. Instead, volatiles will slowly be released from the volatile sublimation fronts, diffuse to the surface, and ultimately leave the body. The rate of volatile loss (and thus rate of sublimation front recession) would depend on the relative volatility of the species. For Example, $CO_2$'s high volatility would cause this species to recede faster into the surface than $H_2O$. Eventually, these differences would be expected to produce a mature equilibrium subsurface structure on Arrokoth, similar to the expected subsurface structures of comets (*Prialnik et al. 2004*). Although these volatile fronts would continue to recede into Arrokoth, their extremely low outgassing rates (figure 1) suggests that this structure has changed little since formation (volatile fronts moved on the order of centimeters over the age of the solar system). Ultimately, the state observed by the *New Horizons* spacecraft is likely representative of the sublimative state and structure of Arrokoth at the dawn of the Quiescent Period, regardless of whether such unobservable layers exist.

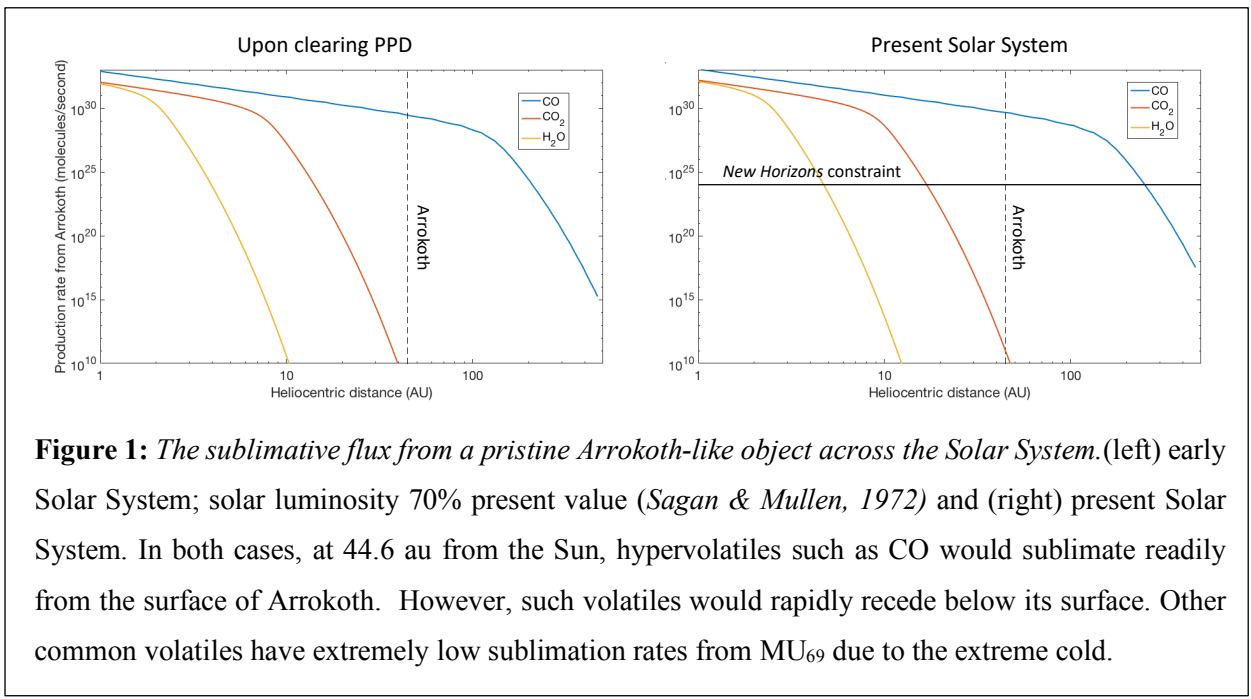

**Figure 1:** *The sublimative flux from a pristine Arrokoth-like object across the Solar System.*(left) early Solar System; solar luminosity 70% present value (*Sagan & Mullen, 1972*) and (right) present Solar System. In both cases, at 44.6 au from the Sun, hypervolatiles such as CO would sublimate readily from the surface of Arrokoth. However, such volatiles would rapidly recede below its surface. Other common volatiles have extremely low sublimation rates from $MU_{69}$ due to the extreme cold.

### III.    Sublimative Spin-State Evolution

Sublimative outgassing of volatiles can produce torques that change an object's rotation state. Since Arrokoth is thought to have formed from the accretion of two lobes to form the bilobate shape observed by the *New Horizons* spacecraft, sublimative torques may have played a role in spinning down the lobes to facilitate merger (*McKinnon et al. 2020*) or spinning up the object to separate the lobes (*Hirabayashi et al. 2020*). However, the effects of sublimation early in Arrokoth's history (e.g., immediately after the clearing of the PPD) are much stronger than later in Arrokoth's history, due to supervolatiles being readily available near its surface. This requires that we explore these two regimes separately. Here we explore the limits and effects of such sublimative torques on the rotation state of Arrokoth.

To compute how such outgassing may change the rotation rate of Arrokoth, we use the SYORP sublimative torque model (*Steckloff & Jacobson, 2016; Steckloff & Samarasinha, 2018*), which repurposes the mathematical machinery of the YORP effect (taking advantage of the similarities in the behavior of radiative and sublimative processes). From this model, the maximum sublimative angular acceleration of Arrokoth is

$$\left|\frac{d\omega}{dt}\right| = \frac{3P_S C_S}{4\pi\rho R^2} \tag{6}$$

$$= \frac{3\Phi_{sub} m v_{thermal} C_S}{4\pi\rho R^2} \tag{7}$$

where $P_S$ is the dynamic sublimation pressure exerted by the escaping gases on the surface, $C_S$ is an empirical factor analogous to YORP coefficients, $\rho$ is the bulk density of Arrokoth; $R$ is the mean radius of Arrokoth, approximately 8.5 km (*Stern et al. 2019*), $\Phi_{sub}$ is the sublimative gas flux from the surface, $m$ is the molecular mass of the volatile, and $v_{thermal}$ is the mean molecular thermal velocity of the molecule. This model assumes that the object has a very low albedo (consistent with the 6.3% bond albedo of Arrokoth *[Hofgartner et al. 2020]*)*,* and is composed entirely of a single volatile ice (for example, CO). This model also assumes that the multiple different chemical species composing Arrokoth do not interact with one another. Such interactions can change molecular binding energies, and thus latent heats of sublimation. However, such effects are poorly understood, and beyond the current state of the art of comet science. Note that this equation only provides the magnitude of the angular acceleration, and that sublimative torques could spin the object up or down.

Like the YORP coefficients (*Scheeres, 2007; Rozitis & Green, 2013*), the SYORP coefficient $C_S$ parameterizes how a body's spin pole obliquity and irregular shape leads to net torques (*Steckloff & Jacobson, 2016*). Unlike radiative processes in which the entire surface emits photons, much of the surfaces of icy bodies are sublimatively inactive. Thus, the SYORP coefficient $C_S$ also accounts for the effects of volatile distribution on sublimative torques (*Steckloff & Samarasinha, 2018*). Additionally, SYORP coefficients must account for the collisional nature of gas molecules near the surfaces of icy bodies, which will cause gas molecules to spread out and collide with the surface, weakening the sublimative generation of torques. Pristine icy objects may have SYORP coefficients on the order of ~$0.0001 - 0.001$ (*Steckloff & Jacobson, 2016*), which is an order of magnitude lower than the typical range for YORP coefficients (*Scheeres, 2007; Rozitis & Green, 2013*). However, evolved objects with only partially sublimatively active surfaces likely have SYORP coefficients that are lower than this range.

To more accurately model sublimative torques, we use observed rotation state changes in short-period comets to compute the SYORP coefficients for these "real", evolved objects. We rearrange equation 6 to solve for $C_S$ from observed rotation state changes and nucleus material properties. In cases where the bulk density of the nucleus is not known, we assume a typical range of comet bulk densities of $300 - 700$ kg/m$^3$ (*Steckloff & Samarasinha, 2018*). We compute the SYORP coefficients for comets 2P/Encke, 9P/Tempel 1, 10P/Tempel 2, 19P/Borrelly, 67P/Churyumov-Gerasimenko, and 103P/Hartley 2, and find that they span a relatively narrow range of values, $1.22$ x$10^{-5}$ - $8.43$ x$10^{-5}$ (see Table 1).

We use these values to compute the effects of CO sublimative torques on Arrokoth shortly after formation (see figure 2), with the timescale of Arrokoth's changing rotation period depending

| | Dominant Volatile species | Effective Radius (km) | Bulk Density (kg/m³) | Rotation period (hr) | Observed Change in Rotation Period (min) | Computed SYORP Coefficient ($C_S$) |
|---|---|---|---|---|---|---|
| 2P/Encke | $H_2O$ | 2.4[a][b] | 300-700(assumed)[c] | 11[a][b] | 4[a][b] | $(3.7 - 8.6)$ x$10^{-5}$ |
| 9P/Tempel 1 | $H_2O$ | 2.83[a][b] | $470^{+780}_{-230}$[d][e] | 41[a][b] | 14[a][b] | $1.22^{+2.03}_{-0.60}$ x$10^{-5}$ |
| 10P/Tempel 2 | $H_2O$ | 6.0[a][b] | 300-700(assumed)[c] | 9[a][b] | 0.27[a][b] | $(1.4 - 3.3)$ x$10^{-5}$ |
| 19P/Borrelly | $H_2O$ | 2.52[b] | 560±270[f] | 28[b] | 40[b] | $7.27(\pm 3.50)$ x$10^{-5}$ |
| 67P/Churyumov-Gerasimenko | $H_2O$ | 1.65[b] | 533±6[g] | 12[b] | 21[b] | $8.43(\pm 0.09)$ x$10^{-5}$ |
| 103P/Hartley 2 | $CO_2$ | 0.58[a][b] | 300±100[h] | 18[a][b] | 150[a][b] | $1.70(\pm 0.57)$x$10^{-5}$ |

**Table 1:** *Observed parameters of Jupiter-Family comets, used to compute SYORP coefficients.* The SYORP coefficient $C_S$ is a parameter that can be computed from observations and properties of comet nuclei. We find that the SYORP coefficients are 1-2 orders of magnitude smaller than those for pristine icy objects (*Steckloff & Jacobson, 2016*), and 2-3 orders of magnitude smaller than the analogous YORP coefficients for radiative torques (*Scheeres, 2007; Rozitis & Green, 2013*).

References: [a] Samarasinha & Mueller (2013), [b] Mueller & Samarasinha (2018), [c] reasonable range of cometary bulk densities (Steckloff & Samarasinha, 2018), [d] Richardson et al. (2007), [e] Thomas et al. (2013), [f] Farnham & Cochran (2002), [g] Päzold et al. (2016), [h] Thomas et al. (2013)

on bulk density and efficiency of the sublimative torques (i.e., the SYORP coefficient $C_S$). We assume a bulk density for Arrokoth of between $300 - 500$ kg/m³ (*Hirabayashi et al. 2020; consistent with McKinnon et al. 2020*). The higher end of this plausible density range (500 kg/m³; *Hirabayashi et al. 2020; McKinnon et al. 2020*) is typical of Jupiter Family Comets such as 67P/Churyumov-Gerasimenko (*Pätzold et al. 2016*) or (D/1993 F2) Shoemaker-Levy 9 (*Asphaug & Benz, 1996*). At this density, Arrokoth's lobes would separate at a spin period of 12.5 hours (*Hirabayashi et al. 2020*), and sublimative torques would require $\sim 0.1 - 10$ Myr to spin down Arrokoth to its present 15.92 hour spin period (assuming its maximum outgassing rate). However, such an outgassing rate would deplete Arrokoth of CO in only $\sim 10,000$ years. As a result, sublimative torques would have changed Arrokoth's rotation period by only as many as a few minutes to tens of minutes (depending on density and SYORP coefficient).

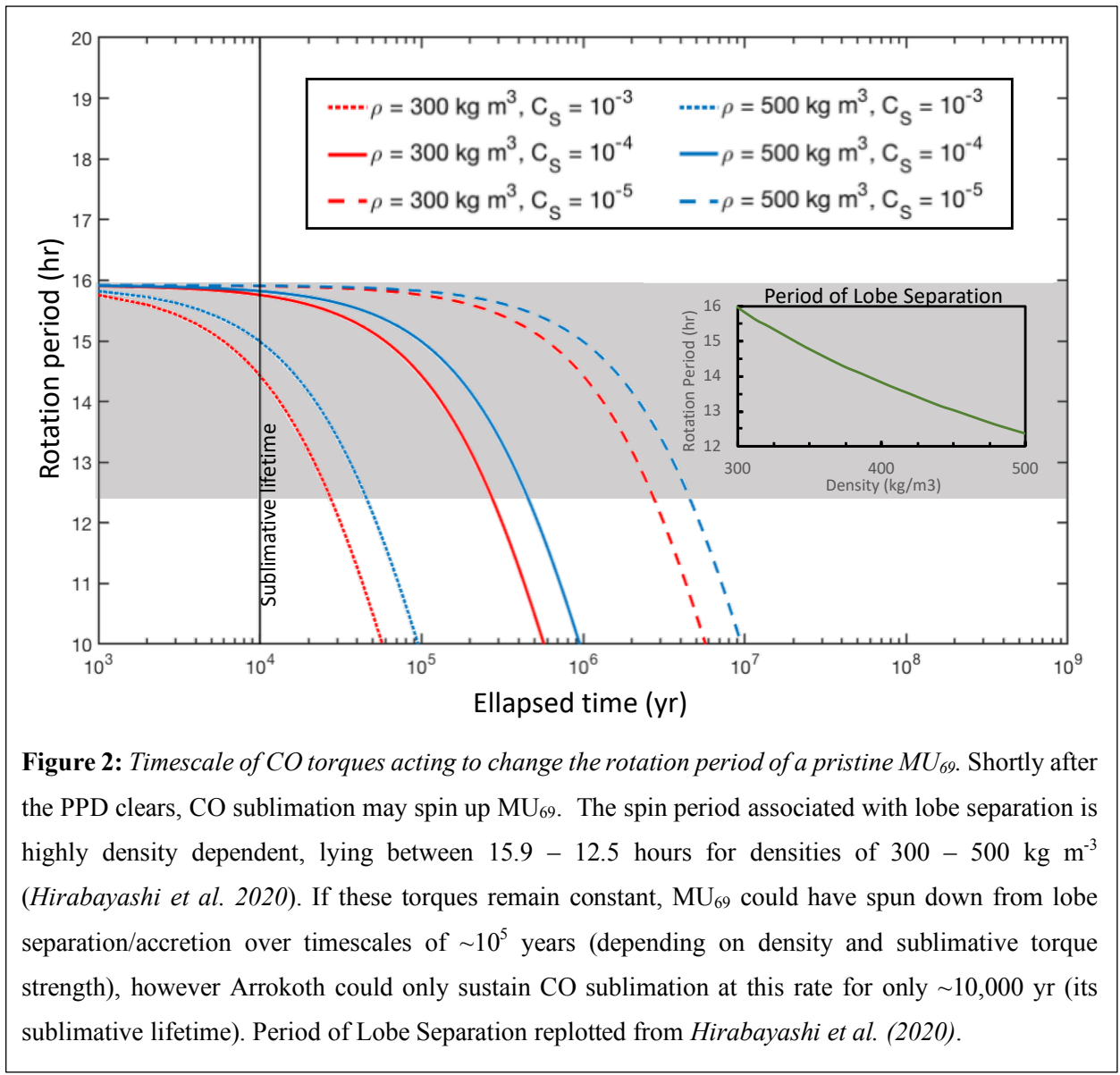

**Figure 2:** *Timescale of CO torques acting to change the rotation period of a pristine $MU_{69}$.* Shortly after the PPD clears, CO sublimation may spin up $MU_{69}$. The spin period associated with lobe separation is highly density dependent, lying between $15.9 - 12.5$ hours for densities of $300 - 500$ kg m⁻³ (*Hirabayashi et al. 2020*). If these torques remain constant, $MU_{69}$ could have spun down from lobe separation/accretion over timescales of $\sim 10^5$ years (depending on density and sublimative torque strength), however Arrokoth could only sustain CO sublimation at this rate for only $\sim 10,000$ yr (its sublimative lifetime). Period of Lobe Separation replotted from *Hirabayashi et al. (2020).*

Although these timescales imagine a maximally outgassing Arrokoth, in reality such volatile loss rates would slow down as the sublimation front receded below the surface, allowing outgassing to last until the solar thermal wave reaches Arrokoth's center (the entire length of the sublimative period). Nevertheless, our modeling effort depends solely on the total volatile flux; the outgassing rate and duration are irrelevant over secular timescales so long as the total volatile flux is the same.

Arrokoth's shape is minimally stable at the lower end of its range of plausible densities of $\sim 300 - 500$ kg/m$^3$ (*Hirabayashi et al. 2020; McKinnon et al. 2020*), a density consistent with the density of comet 103P/Hartley 2 (*Thomas et al. 2013; Richardson & Bowling, 2014*). At such densities, Arrokoth's lobes separate at spin periods just below its present-day spin period; and sublimative torques could have caused the lobes to separate. However, at densities of even a few $\sim 10$'s of kg/m$^3$ above this lower limit, sublimative torques would be insufficient to change Arrokoth's spin state sufficiently to cause its lobes to separate. Ultimately, sublimative torques are unlikely to have a significant effect on Arrokoth's spin state during the Sublimative Period, changing its spin period by a few minutes at most, before vigorous sublimation ceases.

Outgassing during Arrokoth's Quiescent Period similarly had a negligible effect on Arrokoth's spin state. We rearrange equation 7 to solve for the flux required to spin the nucleus up/down

$$\Phi_{sub} = \frac{4\pi\rho R^2 \Delta\omega}{3 m v_{thermal} C_S \Delta t} \tag{8}$$

where $\Delta t$ is the duration of volatile sublimation (here 4.5 billion years) and $\Delta\omega$ is the change in angular speed, which we set to $10^{-6}$ to represent spin period changes on the order of a few minutes (comparable to the limit in the Sublimative Period). We use best-case parameters for Arrokoth to compute the lowest sublimative flux needed for this change: low density of 300 kg/m$^3$ (*Hirabayashi et al. 2020; McKinnon et al. 2020*), peak temperature of 58 K (*Grundy et al. 2020*) across the body, and high SYORP coefficient of $10^{-4}$. Even with these best-case values, the average $CO_2$ flux from Arrokoth would need to be $\sim 10^{17}$ molecules/s to change its spin period by a few minutes. However, the sublimative flux of $CO_2$, the most volatile major component of cometary ices following CO, would have a maximum flux from Arrokoth of only $\sim 10^{11}$ molecules/s (figure 1). For lower SYORP values and higher bulk densities (which are more consistent with other typical Cold Classical Kuiper Belt Object density estimates of ~600 – 1400 kg/m³; *Vilenius et al. 2012; Grundy et al. 2012; Brown, 2013; Vilenius et al. 2014*), the minimum flux increases to $\sim 10^{18}$ molecules/s.  Other, more volatile ices may contribute to this flux, but would have minor contributions.  Thus, Arrokoth's spin state has changed negligibly since the end of the Sublimative Period.

Finally, we have thus far treated sublimative torques as monotonic increases or decreases in the rotation state of an object.  However, sublimative torques are likely stochastic in direction, both spinning objects up and down (*Statler, 2009; Cotto-Figueroa et al. 2015)*.  With this consideration, the calculated changes in rotation rates may overestimate actual changes by an order of magnitude.

Considering that impacts have also changed Arrokoth's spin period by at most a few minutes (*Hirabayashi et al. 2020*), the present spin period is representative of Arrokoth's spin state upon the clearing of the Sun's protoplanetary disk.  Similarly, sublimative torques are unlikely to have contributed to the initial formation of Arrokoth's contact binary shape.  Arrokoth's shape is indicative of a gentle merger of the two lobes, suggesting a gentle process for eliminating orbital angular momentum from the system to cause the lobes to merge (*McKinnon et al. 2020*), such as nebular drag (*McKinnon et al. 2020; Lyra et al. 2020 submitted*).  Sublimative torques can only change the present angular momentum of the Arrokoth system by less than ~1% (assuming ideal conditions), and therefore contributed negligibly to the initial merger of Arrokoth's lobes.  This further supports the notion that Arrokoth's shape is primordial (*McKinnon et al. 2020*).

## IV.  Sublimative Surface Evolution

Volatile sublimation can drive topographic erosion (*Britt et al. 2004; Steckloff & Samarasinha, 2018*) and mass transport about the surface of an icy body (*Lai et al. 2016*).  Rates of sublimation-driven erosion/deposition likely vary significantly within classes of icy bodies, and between different dynamical classes of bodies.  Although a full calculation of these erosion/deposition rates on Arrokoth would require detailed simulations with many underconstrained variables.  Nevertheless, if we assume that sublimation-driven evolution of Jupiter Family Comets is representative (when scaled) of icy bodies in the solar system, we can use existing studies of mass transport on JFCs to estimate the magnitude of mass transport/deposition on Arrokoth.

During the sublimative era, the surface vigorously sublimated CO and other supervolatiles. Using the sublimation model of *Steckloff et al. (2015)*, we find that, even though CO is significantly more volatile than $H_2O$ ice, the dynamic sublimation pressure driving and transporting dust about Arrokoth's surface is only ~0.5% as strong as $H_2O$'s dynamic sublimation pressure on a typical JFC. Additionally, Arrokoth's large size gives it a surface escape speed and surface gravity on the order of 5 times greater than those of a typical JFC such as 67P/Churyumov-Gerasimenko (assuming comparable densities). Comparing these weak drag forces and higher gravity, the maximum loftable dust grain from Arrokoth would be ~0.1% the radius of the maximum loftable grain on 67P/Churyumov-Gerasimenko (on the order of hundreds of microns, rather than on the order of decimeters; *Lai et al. 2016; Keller et al. 2017*).

The grain size frequency distribution modeled in *Lai et al. (2016)* has a break in the power law differential size-frequency distribution at around a millimeter, with grains smaller than this following a power-law index of 2, and grains larger than this following a power-law index of 4 (a steeper drop-off with size (a power-law index of 3 indicates that mass is equally distributed amongst the grain sizes). Therefore, less mass is in the population of larger grains than in the population of smaller grains. Thus, if a similar grain-size distribution held true on Arrokoth, then the loss of these larger grains (which cannot be lofted on Arrokoth) would reduce the transported mass per unit area by less than 0.1% (relative to 67P). We adopt a 0.1% value as a conservative estimate.

Numerical models of thermally driven erosion and deposition are highly sensitive to model parameters. Nevertheless, numerous models explored the role of such processes on comet 67P/Churyumov-Gerasimenko, a Jupiter-Family Comet that was the target of the *Rosetta* mission, and found broad agreement in erosion/deposition rates. Global models found that the surface of 67P/Churyumov-Gerasimenko can erode on the order of ~4 m/year (*Lai et al. 2016; Keller et al. 2017*), and experience depositional growth on the order of ~0.5 m/yr (*Lai et al. 2016*) during an apparition. This is consistent with analysis of *Rosetta* images, which suggest erosion in many region of ~1 m and deposition of around ~20 cm (*Xu et al. 2017*). Other modeling efforts find that smooth terrains on 67P/Churyumov-Gerasimenko experience topographic evolution of ~1 m or less (*Tang et al. 2019*) or ~20 cm (*Kossacki & Jasiak, 2019*), although these terrains are themselves depositional, and can experience subsequent thermally driven topographic evolution (*Birch et al.*

*2019*). These results are consistent with the expectation that topographic changes occur more rapidly on steeper terrains than flatter terrains (*Steckloff et al. 2018*).

Ultimately, these studies suggest that the erosion rate on comet 67P/Churyumov-Gerasimenko is on the order of ~0.2 – 4 m/year; thus, the scaled erosion rate on Arrokoth would be ~0.02 - 0.4 cm/yr. Arrokoth's higher escape speed and weaker sublimation pressures would result in negligible grain escape (unlike Jupiter family comets), resulting in a deposition rate that effectively matches the erosion rate. Thus, over the sublimative lifetime of Arrokoth (the time in which a block of CO the size of Arrokoth would completely disappear) of ~10,000 yr, the topography may erode/fill in, and ultimately change by as much as ~10 - 100 m during the sublimative period.

By comparison, sublimative forces during the quiescent period are effectively non-existent. The outgassing constraint observed by the *New Horizons* spacecraft would result in sublimation pressures more than 13 orders of magnitude smaller than those during the Sublimative Period. Dust transport under such conditions would undoubtedly be negligible, consistent with the lack of an observable dust coma during the *New Horizons* flyby.

These mass transport processes could profoundly alter the primordial surface of Arrokoth. Such erosion and deposition could effectively erase craters up to ~50 – 500 m across (assuming canonical depth/diameter ratio of 0.2). Indeed, most of the observed and suspected craters on Arrokoth fall in the upper end of this size range (*Spencer et al. 2020*). This suggests, rather provocatively, that the majority of the observed surface morphology and craters on Arrokoth may not date not from the Sublimative Period, but rather from the Quiescent Period. In other words, although Arrokoth's shape is likely primordial, its surface may not be.

### V.    Conclusions

We consider the effects of sublimating volatiles on (486958) Arrokoth, and find that Arrokoth's sublimative history can be separated into two periods. An early Sublimative Period lasting ~10-100 Myr after the clearing of the Sun's protoplanetary disk (PPD), during which Arrokoth's supervolatile species such as CO sublimated vigorously. Such sublimation would cause significant erosion and dust transport about the surface, erasing surface features shallower than ~10 - 100 m. This includes craters up to ~50 – 500 m across, suggesting that the surface of Arrokoth we see today may not be primordial, but rather could date from after the Sublimative Period. Arrokoth's Sublimative Period was followed by a Quiescent Period, which continues to the present day. The Quiescent Period is marked by a lack of significant outgassing, with expected volatile production rates at least 13 orders of magnitude smaller than what *New Horizons* could detect. Sublimative torques have been insignificant throughout Arrokoth's history, suggesting that the present spin rate is representative of Arrokoth's primordial spin state, and that Arrokoth's shape, unlike its surface, is indeed primordial.

## VI.    Acknowledgments

The authors with to thank Seth Jacobson for productive conversations that helped explore the implications of this work.  The authors also with to thank both anonymous reviewers, whose comments greatly improved the content, structure, organization, and ultimately, the conclusions and implications of this work. J.K.S. and G.S. acknowledge support from NASA award 80NSSC18K0497.